\documentclass[fleqn,10pt]{wlscirep1}
\usepackage{changes}
\usepackage{lipsum}% <- For dummy text
\usepackage{multirow}
\usepackage{hyperref}
\usepackage{fix-cm}
\usepackage{amsmath}
\usepackage{graphicx}
\usepackage{verbatim}
\usepackage{tabularx}
\usepackage{ctable}
\usepackage{array}
\usepackage{xcolor}
\usepackage{colortbl}
\usepackage{hhline}
\usepackage{xr}
\usepackage{caption}
\usepackage{graphics}
\usepackage{epsfig}

\usepackage{etoolbox}
\makeatletter
\patchcmd{\@maketitle}{ABSTRACT}{}{}{}
\makeatother
\usepackage{sidecap}

\title{Network approach integrates 3D structural and sequence data to improve protein structural comparison}

\author[1,5,6]{Fazle E. Faisal}
\author[2,7]{Julie L. Chaney}
\author[1,5,6]{Khalique Newaz}
\author[3]{Jun Li}
\author[1]{Scott J. Emrich}
\author[2,4,6]{Patricia L. Clark}
\author[1,5,6,*]{Tijana Milenkovi\'{c}}
\affil[1]{Department of Computer Science and Engineering, University of Notre Dame, Notre Dame, IN 46556, USA}
\affil[2]{Department of Chemistry and Biochemistry, University of Notre Dame, Notre Dame, IN 46556, USA}
\affil[3]{Department of Applied and Computational Mathematics and Statistics, University of Notre Dame, Notre Dame, IN 46556, USA}
\affil[4]{Department of Chemical and Biomolecular Engineering, University of Notre Dame, Notre Dame, IN 46556, USA}
\affil[5]{Interdisciplinary Center for Network Science and Applications, University of Notre Dame, Notre Dame, IN 46556, USA}
\affil[6]{ECK institute for Global Health, University of Notre Dame, Notre Dame, IN 46556, USA}
\affil[7]{Siemens Healthineers, Elkhart, IN 46516, USA}

\affil[*]{Corresponding author (email: tmilenko@nd.edu)}

\begin{abstract}
Initial protein structural comparisons were sequence-based.  Since
amino acids that are distant in the sequence can be close in the
3-dimensional (3D) structure, 3D contact approaches can complement
sequence approaches.  Traditional 3D contact approaches study 3D
structures directly. Instead, 3D structures can be modeled as protein
structure networks (PSNs). Then, network approaches can compare
proteins by comparing their PSNs. Network approaches may improve upon
traditional 3D contact approaches. We cannot use existing PSN
approaches to test this, because: 1) They rely on naive measures of
network topology. 2) They are not robust to PSN size. They cannot
integrate 3) multiple PSN measures or 4) PSN data with sequence data,
although this could help because the different data types capture
complementary biological knowledge. We address these limitations by:
1) exploiting well-established graphlet measures via a new network
approach, 2) introducing normalized graphlet measures to remove the
bias of PSN size, 3) allowing for integrating multiple PSN measures,
and 4) using ordered graphlets to combine the complementary PSN data
and sequence data. We compare both synthetic networks and real-world
PSNs more accurately and faster than existing network, 3D contact, or
sequence approaches. Our approach finds PSN patterns that may be
biochemically interesting.
\end{abstract}
\begin{document}

\flushbottom
\maketitle

\thispagestyle{empty}

\section{Introduction}

Proteins perform important cellular functions. While understanding
protein function is clearly important, doing so experimentally is
expensive and time-consuming \cite{Ashburner2000,Kasabov2013}. Because
of this, the functions of many proteins remain unknown
\cite{Lee2007,Kasabov2013}. Consequently, computational prediction of
protein function has received attention. In this context, protein
structural comparison (PC) aims to quantify similarity between
proteins with respect to their sequence or 3-dimensional (3D)
structural patterns, in order to predict functions of unannotated
proteins based on functions of annotated proteins that they are
similar to.

Early PC has relied on sequence analyses
\cite{Mizianty2010,Sulkowska2012}. Due to advancements of
high-throughput sequencing technologies, rich sequence data is
available for many species, and thus, comprehensive sequence pattern
searches are possible.

Amino acids that are distant in the linear sequence can be close in 3D
structure. Thus, 3D structural analyses can reveal patterns that might
not be apparent from the sequence alone \cite{Kihara2005}. For example, while high
sequence similarity between proteins typically indicates their high
structural and functional similarity \cite{Lee2007}, proteins with low
sequence similarity can still be structurally similar and perform
similar function \cite{Krissinel2006,Gao2009}. In this case, 3D
structural approaches, unlike sequence approaches, can correctly
identify structurally and thus functionally similar proteins.  On the
other extreme, proteins with high sequence similarity can be
structurally dissimilar and perform different functions
\cite{Tuinstra2008,Kosloff2008,Clarke2008,Burmann2012}. In this case,
3D structural approaches, unlike sequence approaches, can correctly
identify structurally and thus functionally different proteins.

3D structural approaches can be categorized into traditional 3D
contact approaches and recent network approaches. 3D contact
approaches, which typically deal with 3D structural alignment, study
3D structures directly \cite{DaliLite2010,TMAlign2005}.  Instead,
network approaches first model 3D structures as protein structure
networks (PSNs, or contact maps, in which nodes are amino acids and
edges link spatially close amino acids \cite{Milenkovic2009null}) and
then compare proteins by comparing their PSNs. 3D contact approaches
produce rigid protein alignments while comparing 3D structures
\cite{Malod2014}. Hence, they may not always perform well in the task of 
PC. Also, 3D contact approaches are typically slow.  This requires
alternative, more flexible and faster PC approaches.  Since PSNs model
spatial proximities of amino acids within the protein 3D structure,
network analyses of PSNs have a potential to complement and improve
upon 3D contact (as well as sequence) approaches in the task of PC.

Network analyses of 3D structures \emph{have} received attention
\cite{Gao2009,Emerson2013}, e.g., when comparing functionally different proteins \cite{Pabuwal2008,Pabuwal2009,Emerson2012}. However, these existing network approaches have limitations:
\vspace{-0.225cm}
\begin{enumerate}
\item They  rely on naive measures of network topology, which capture  
the global view of a network but ignore complex local
interconnectivities that exist in real-world networks, including PSNs
\cite{Milenkovic2008gc,Memisevic2010b,Kuchaiev2011}.  Hence, a more 
sensitive measure of local network topology might improve PC.
\vspace{-0.225cm}
\item They can bias PC by PSN size: 
networks of similar topology but different sizes can be mistakely
identified as dissimilar by the existing approaches simply because of
their size difference. Thus, PC strategies are needed that can avoid
the PSN size bias.
\vspace{-0.225cm}
\item Because different network measures quantify the same PSN topology 
from different perspectives \cite{Emerson2013,Faisal2014dynamic}, and
because the existing approaches rely on a single network measure, PC
could be biased towards the PSN perspective captured by the given
measure. Integration of different and complementary network measures
could improve PC.
\vspace{-0.225cm}
\item Almost all existing network approaches ignore valuable sequence 
information (also, the existing sequence approaches ignore valuable
PSN information). Combining the complementary ideas of network and
sequence measures could improve PC.
\end{enumerate}

\vspace{-0.225cm}

We present a new network-based PC framework that relies on principal
component analysis (PCA) and that overcomes the above drawbacks of the
existing approaches. Specifically:
\vspace{-0.225cm}
\begin{enumerate}
\item We use graphlets \cite{Przulj2004,Przulj2007}, a sensitive measure 
of \emph{local} network topology, in hope to improve PC upon the
existing network approaches. While graphlets have already been proven
in analyses of protein-protein interaction networks
\cite{Hulovatyy2014,Hulovatyy2015,Solava2012}, here we use them in a
novel application of PC. Also, we use them within our PCA framework,
where none of the existing graphlet methods rely on PCA.
\vspace{-0.225cm}
\item We perform graphlet normalization to 
address the bias of PSN size.
\vspace{-0.225cm}
\item We allow for integrating different and complementary network 
topological measures within our framework.
\vspace{-0.225cm}
\item We adopt the idea of \emph{ordered graphlets} \cite{Malod2014} to 
integrate the PSN amino acid interconnectivity information with
sequence information, in order to improve upon network approaches
alone or sequence approaches alone.  While ordered graphlets are an
existing idea, this idea was introduced only on up to 3-node
graphlets. However, using larger graphlets can be beneficial in many
real-world contexts
\cite{Faisal2014dynamic,Hulovatyy2014,Hulovatyy2015,Solava2012}. 
Hence, we extend the existing notion of 3-node ordered graphlets both
theoretically and implementation-wise to be able to deal with larger
graphlets. Importantly, even when we use only the existing up to
3-node ordered graphlets, our new PCA framework already outperforms
the existing approach that is based on the same ordered graphlets
\cite{Malod2014}. This validates the PCA framework as a
whole. Using larger graphlets helps further. Additionally, we
introduce a novel concept of
\emph{``long-range($K$)'' ordered graphlets} to give higher importance to 
amino acids that are close enough in the protein 3D structure but are
at least $K$ amino acids apart in the protein sequence than to amino
acids that are close enough in the 3D structure simply because they
are also close to each other in the sequence, as such longer-range
interactions might help distinguish protein structures better
\cite{Gromiha2004,Kihara2005}. Indeed, ``long-range($K$)'' ordered graphlets further 
improve accuracy compared to traditional ordered graphlets.  We
include the implementation of the extended idea of (larger size as
well as ``long-range($K$)'') ordered graphlets into our software (available upon request).
\end{enumerate}

\vspace{-0.225cm}

\begin{figure*}[!t]
\centering
\includegraphics[scale=0.15]{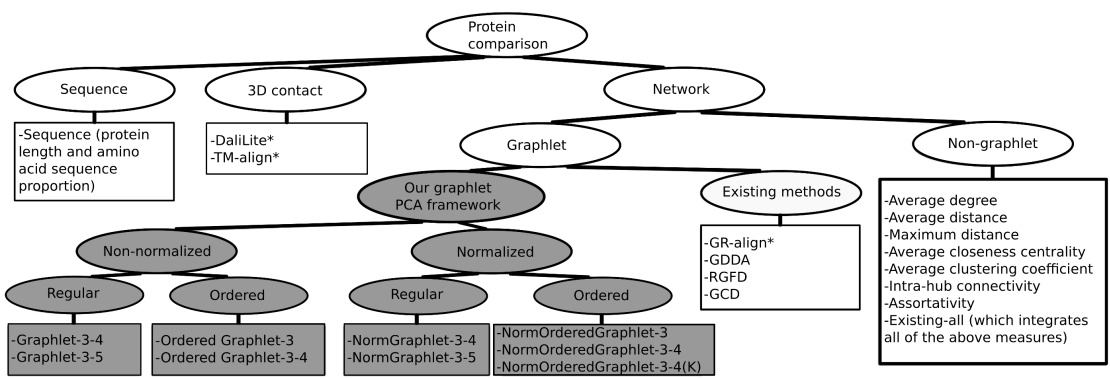}
\caption{Categorization of the 24 approaches (listed in squares) that we evaluate.  Different versions of our proposed graphlet PCA approach are colored in grey. Alignment-based approaches are marked with an asterisk (*) sign; all remaining approaches are alignment-free (see
Methods for details).}
\label{fig:dendogram}
\end{figure*}

We study two network types: synthetic networks (in order to illustrate
wide applicability of our approach across many domains) and real-world
PSNs (in order to illustrate a specific application of our approach in
the task of PC). For each network type, we analyze multiple data
sets. In each data set, each network has a known label, meaning that
we know that networks having the same label should be identified as
similar, while networks having different labels should be identified
as dissimilar.  For synthetic networks, we study 21 network approaches
(Fig. \ref{fig:dendogram}), of which nine are different versions of
our proposed graphlet PCA approach and 12 are existing (non-PCA)
approaches (of which four use graphlets and eight do not use
graphlets).  Given a data set and a network approach, we compute
similarity/distance between each pair of networks.  We evaluate each
approach by measuring how accurately it can identify as similar
networks of the same label and as dissimilar networks of different
labels. We measure this by computing the area under precision-recall
curve (AUPR) and area under receiver operator characteristic curve
(AUROC).  For real-world networks, in addition to the 21 network
approaches, we also study two 3D contact approaches and a sequence
approach, and we perform the same AUPR and AUROC evaluation. (These
three approaches
\emph{cannot} be used on the synthetic networks.) For details, see Methods.

Our key findings are as follows. Over all data sets, our graphlet PCA
approach is superior to the existing (non-PCA) graphlet and
non-graphlet network approaches. We demonstrate the importance of
removing the bias in PSN size, leading to a normalized version of our
graphlet PCA approach that is superior to its non-normalized
counterpart when controlling for network size.  Combining our
normalized graphlet PCA approach with sequence data via ordered
graphlets results in superior accuracy compared to any network
approach alone or sequence approach alone, which confirms that data
integration helps. Adding the ``long-range($K$)'' constraint on the
normalized ordered graphlets further improves accuracy, which
additionally confirms the importance of long-range amino acid
interactions. Our network approach outperforms the traditional 3D
contact approaches in terms of both accuracy and running time, which
confirms the power of network analyses of 3D structures. Our approach
reveals PSN patterns that may be biochemically interesting. For
details, see Results.

\section{Methods}
\subsection{Data}
\label{method:data}

We collect 3D atomic structures of proteins from the Protein Data Bank
(PDB) \cite{PDB}, where each protein is annotated by a label from
CATH, SCOP, or both (see below). Since PDB contains multiple copies of
the same or nearly identical proteins, we aim to reduce the redundancy
by selecting a set of proteins from PDB such that each protein in the
set is not more than 90\% sequence identical to any other protein in
the set. If a protein is not more than 90\% sequence identical to any
other protein from PDB, we immediately select the protein. If a
protein is more than 90\% sequence identical to one or more proteins
from PDB, we select a ``representative'' protein from such a protein
group so that the representative protein is of the highest quality (in
terms of resolution) among all proteins in the group.  This strategy
results in the selection of 17,036 proteins. We denote this data set
as \emph{ProteinPDB}. Each protein in the data is comprised of the X,
Y, and Z orthogonal Angstrom (\r{A}) coordinates of heavy atoms (i.e.,
\emph{carbon}, \emph{nitrogen}, \emph{oxygen}, and \emph{sulfur}) of
each amino acid within the protein. The data is available at
http://www.rcsb.org/pdb/home/home.do for free download.

Class, Architecture, Topology, Homology (CATH) is a protein domain
categorization database \cite{Sillitoe2015,Orengo1999}. A protein is
typically composed of one or more domains (a domain refers to a common
protein structure), and the purpose of CATH is to annotate these
domains. CATH's categorization scheme is hierarchical. Its top
hierarchy divides protein domains into four groups (i.e., categories
or labels):
\emph{alpha} ($\alpha$), \emph{beta} ($\beta$), \emph{alpha/beta}
($\alpha$/$\beta$), and \emph{few secondary structures}. Only for few
secondary structures, none of the domains in ProteinPDB belong to this
category, and so we remove few secondary structures from further
consideration. Each of the remaining three top-level CATH categories
has deeper-level subcategories, which we also consider, per our
discussion below.

Structural Classification of Proteins (SCOP) \cite{Murzin1995} is
another protein domain categorization database whose categorization
scheme is also hierarchical. SCOP's top hierarchy divides protein
domains into 11 groups: $\alpha$, $\beta$, $\alpha$/$\beta$,
\emph{alpha plus beta} ($\alpha$+$\beta$), \emph{coiled coil},  
\emph{membrane}, \emph{multi-domain}, \emph{small}, \emph{low resolution}, 
\emph{peptide}, and \emph{designed}. Only for small, low resolution, 
peptide, or designed, none of the domains in ProteinPDB belong to
these categories, and so we remove these four categories from further
consideration.  Each of the remaining top-level SCOP categories has
deeper-level subcategories, which we also consider, per our discussion
below.

\subsection{Forming networks}
\label{method:network}
We evaluate the considered approaches in the task of PC on: 1)
synthetic networks, i.e., on artificially generated networks for which
we know the topology-based ground truth categorization, and 2)
real-world PSNs, for which we know CATH or SCOP label-based
categorizations that we hypothesize correlate well with the PSNs'
topology-based characteristics.

\vspace{0.1cm}

\noindent\textbf{Synthetic networks.}
We generate synthetic networks by using different network models.  A
good approach should identify networks from the same network model
(i.e., with the same label) as similar, and it should identify
networks from different models (i.e., having different labels) as
dissimilar. Specifically, we use three well-established network
models: \emph{Erd\H{o}s-R\'{e}nyi random graphs (ER)}, \emph{geometric
random graphs (GEO)}, and \emph{scale-free random graphs (SF)}
\cite{Milenkovic2008gc,Kuchaiev2011}. We note that these models are
not necessarily representative of PSNs.  Instead, they are
general-purpose models. This is intentional, because the models that
we use are intended to illustrate wide applicability of our approach
to any domain where data can be modeled as networks. It is our
subsequent analyses on real-world PSNs that will focus specifically on
the task of PC.

First, we evaluate the considered approaches on synthetic networks of
the same size but of different labels (originating from the three
network models). To evaluate the robustness of our PC framework to the
choice of network size, we repeat this analysis three times, by
increasing the size of the considered networks. That is, we perform
three separate analyses of three different network data sets, where in
a given data set, all networks are of the same size, and one third of
the networks in the set comes from each of the three network models.
We denote these network sets as \emph{Synthetic-100},
\emph{Synthetic-500}, and \emph{Synthetic-1000} (Table
\ref{tab:network-data}), where each set consists of 50 networks per model 
(totaling to $50 \times 3 = 150$ networks). The numbers of nodes and
edges in these networks are chosen in a way so that the networks
closely mimic sizes of real-world PSNs.

Second, we evaluate the considered approaches on networks of different
sizes as well as different labels, to check whether the approaches can
correctly identify as similar networks from the same model despite the
networks being of different sizes, as well as that they can correctly
identify as dissimilar networks from different models despite the
networks being of the same size.  To generate a synthetic network set
of different sizes, we combine networks from Synthetic-100,
Synthetic-500, and Synthetic-1000 together. We denote the combined
network set as \emph{Synthetic-all} (Table \ref{tab:network-data}).

\vspace{0.1cm}

\noindent\textbf{Real-world PSNs with CATH categorization.}
Each protein in ProteinPDB (defined above) is composed of 3D
coordinates of the heavy atoms of its amino acids. Given a protein, we
use the 3D coordinate information to compute Euclidean distance
between each pair of amino acids, i.e., between any of their
heavy atoms. Then, we construct a PSN in which nodes represent amino
acids and edges connect pairs of amino acids that are sufficiently
close (i.e. within a given distance cut-off) in the protein's 3D
structure. Again, we emphasize that two amino acids are sufficiently
close if any of the heavy atoms of the first amino acid and any of the
heavy atoms of the second amino acid are within a given distance
cut-off. While effective definitions of contact between amino acids may differ from fold to fold \cite{Yuan2012}, we use the suggested distance cut-off of 4 \r{A}
\cite{Milenkovic2009null}. ProteinPDB contains 17,884 protein domains
that have CATH categorization, which results in 17,884 PSNs. Of these
PSNs, to ensure that PSNs are of reasonable ``confidence'', we focus
for further analyses on those PSNs that meet all of the following
criteria: 1) the given network has more than 100 nodes, 2) the maximum
diameter of the network is more than five, and 3) the network is
composed of a single connected component. This results in 9,509 such
PSNs.

First, we test how well the considered approaches can compare PSNs
belonging to the top hierarchical categories of CATH (i.e., $\alpha$,
$\beta$, and $\alpha$/$\beta$). Of the 9,509 PSNs, 2,628, 3,085, and
3,796 PSNs belong to (i.e., are labeled with) $\alpha$, $\beta$, and
$\alpha$/$\beta$ categories, respectively. We denote this set as
\emph{CATH-primary}. The set contains a large enough number of PSNs in
each category, which ensures enough statistical power for further
analyses.

Second, we test how well the approaches can compare PSNs belonging to
the secondary hierarchical categories of CATH. To ensure enough
statistical power for further analyses, we pick all secondary
categories of $\alpha$, $\beta$, and $\alpha$/$\beta$ that comprise of
at least 30 PSNs. We denote these three sets as \emph{CATH-$\alpha$},
\emph{CATH-$\beta$}, and
\emph{CATH-$\alpha$/$\beta$}, respectively. CATH-$\alpha$ consists of
four secondary $\alpha$ categories (i.e., labels), with an average of
656 PSNs per category. CATH-$\beta$ consists of 10 secondary $\beta$
categories, with an average of 297 PSNs per
category. CATH-$\alpha$/$\beta$ consists of 4 secondary
$\alpha$/$\beta$ categories, with an average of 948 PSNs per
category. For details, see Table \ref{tab:network-data} and Supplementary Table S2.%\ref{supple:tab:secondlevel}.

\vspace{0.1cm}

\noindent\textbf{Real-world PSNs with SCOP categorization.}
ProteinPDB has 15,762 protein domains with SCOP categorization, which
results in 15,762 PSNs. Of these PSNs, to ensure that PSNs are of
reasonable ``confidence'', we focus for further analyses on 11,451
PSNs that meet all of the above criteria while forming PSNs with CATH
categorization.

Per our above strategy (when analyzing PSNs with CATH categorization),
first, we evaluate how well the considered approaches can compare PSNs
from the top hierarchical categories of SCOP (i.e. $\alpha$, $\beta$,
$\alpha$/$\beta$, $\alpha$+$\beta$, coiled coil, membrane, and
multi-domain). Of the 11,451 PSNs, 1,678, 2,541, 3,835, 2,879, 44,
156, and 318 PSNs belong to $\alpha$, $\beta$, $\alpha$/$\beta$,
$\alpha$+$\beta$, coiled coil, membrane, and multi-domain categories,
respectively. This set, denoted as \emph{SCOP-primary}, contains
enough PSNs in each category to ensure enough statistical power for
further analyses.

Second, we evaluate how well the approaches can compare PSNs belonging
to the secondary hierarchical categories of SCOP. To ensure enough
statistical power, we pick all secondary categories of $\alpha$,
$\beta$, $\alpha$/$\beta$, $\alpha$+$\beta$, coiled coil, membrane,
and multi-domain that comprise of at least 30 PSNs. We denote these
five sets as
\emph{SCOP-$\alpha$},
\emph{SCOP-$\beta$}, \emph{SCOP-$\alpha$/$\beta$},
\emph{SCOP-$\alpha$+$\beta$}, and \emph{SCOP-multidomain},
respectively.  SCOP-$\alpha$ consists of 16 secondary $\alpha$
categories, with an average of 57 PSNs per category. SCOP-$\beta$
consists of 21 secondary $\beta$ categories, with an average of 88
PSNs per category.  SCOP-$\alpha$/$\beta$ consists of 26 secondary
$\alpha$/$\beta$ categories, with an average of 113 PSNs per
category. SCOP-$\alpha$+$\beta$ consists of 28 secondary
$\alpha$+$\beta$ categories, with an average of 57 PSNs per category.
SCOP-multidomain consists of 2 secondary multi-domain categories, with
an average of 63 PSNs per category. For details, see Table \ref{tab:network-data} and Supplementary Table S2.%\ref{supple:tab:secondlevel}.

\begin{SCtable}
%\begin{center} 
\caption
{Synthetic network and real-world PSN data sets that we use. For the given data set, the second column indicates whether its networks are of the same size or different sizes, and the last three columns indicate the number of its networks as well as their size(s) in terms of the number of nodes and edges. For more details, see Supplementary Table S2. %\ref{supple:tab:secondlevel}.
}
\fontsize{7}{9}\selectfont
\begin{tabular}{|ll|l|l|ccc|}
\hline
\multicolumn{4}{|c|}{Data set} & \multicolumn{3}{c|}{Number of} \\
\hline
\multicolumn{2}{|l|}{Type} & Size & Name & Networks & Nodes & Edges \\
\hline
\multirow{4}{*}{\rotatebox[origin=c]{90}{Synthetic}} & \multirow{4}{*}{\rotatebox[origin=c]{90}{networks}} & \multirow{3}{*}{Same} & Synthetic-100 & 150 & 100 & 400 \\
& & & Synthetic-500 & 150 & 500 & 2,000 \\ & & & Synthetic-1000 & 150
& 1,000 & 4,000 \\
\cline{3-7}
& & Different & Synthetic-all & 450 & 100-1,000 & 400-4,000 \\
\hline
\multirow{13}{*}{\rotatebox[origin=c]{90}{Real-world}} & \multirow{13}{*}{\rotatebox[origin=c]{90}{PSNs}} & \multirow{3}{*}{Same} & CATH-95 & 24 & 95 & 343-362 \\
& & & CATH-99 & 28 & 99 & 347-374 \\
& & & CATH-251-265 & 16 & 251-265 & 1,003-1,076 \\
\cline{3-7}
& & \multirow{10}{*}{Different} & CATH-primary & 9,509 & 101-872 & 243-3,849 \\
& & & CATH-$\alpha$ & 2,628 & 101-872 & 320-3,849 \\
& & & CATH-$\beta$ & 3,085 & 101-559 & 243-2,166 \\
& & & CATH-$\alpha$/$\beta$ & 3,796 & 101-759 & 288-3,507 \\
\cline{4-7}
& & & SCOP-primary & 11,451 & 101-1,381 & 105-5,558 \\
& & & SCOP-$\alpha$ & 1,678 & 101-938 & 147-4,082 \\
& & & SCOP-$\beta$ & 2,541 & 101-581 & 111-2,113 \\
& & & SCOP-$\alpha$/$\beta$ & 3,835 & 101-904 & 105-3,966 \\
& & & SCOP-$\alpha$+$\beta$ & 2,879 & 101-696 & 120-3,064 \\
& & & SCOP-multidomain & 318 & 196-1,256 & 767-5,558 \\
\hline
\end{tabular}
\label{tab:network-data} 
%\end{center}
\end{SCtable}

\vspace{0.1cm}

\noindent\textbf{Real-world PSNs of the same size.}
To ensure that PC is not biased by PSN size, we need a data set with
PSNs of the same (or similar) network size.  Hence, focusing on PSNs
of $\alpha$ and $\beta$ categories from the CATH-primary data set, we
infer three such same-size PSN data sets, denoted as
\emph{CATH-95}, \emph{CATH-99}, and \emph{CATH-251-265} (Table \ref{tab:network-data} and Supplementary Section S1.1).%\ref{supple:method:network}).

\subsection{Our graphlet PCA framework}
\label{method:framework}

\subsubsection{The PCA framework}
\label{method:framework:classification}

The novelty of our new PCA framework comes from using graphlet-based
measures in the task of PC (Fig.\ref{fig:dendogram}).  Yet, the
framework is generalizable, as it can use any measure(s).  Namely,
given a network data set and a measure of network topology (see
below), we compute one vector per network per measure.  We perform PCA
(a standard dimension reduction technique) on the resulting vectors to
compute principal components for each network. We pick the first $r$
principal components, where the value of $r$ is at least two and as
low as possible so that the $r$ components account for at least 90\%
of variation in the data. For every pair of networks $N_i$ and $N_j$,
we compute their cosine similarity, $s^{cos}(N_i,N_j)$, based on the
networks' first $r$ principal components. We convert the similarity
into distance as $d^{cos}(N_i,N_j) = 1 - s^{cos}(N_i,N_j)$. We use the
PCA-based distances to hypothesize that same-label networks will be
close in the PCA space while networks of different labels will be
distant.
Like most of the network approaches from Fig. \ref{fig:dendogram}, our
approach performs \emph{alignment-free} network comparison, i.e., it
does \emph{not} need to align nodes between the compared networks
before it can quantify their similarity, as
\emph{alignment-based} approaches do
\cite{Yaveroglu2015}.

\subsubsection{Our graphlet measures}
\label{method:framework:feature}

\begin{figure}
\centering
\includegraphics[scale=1]{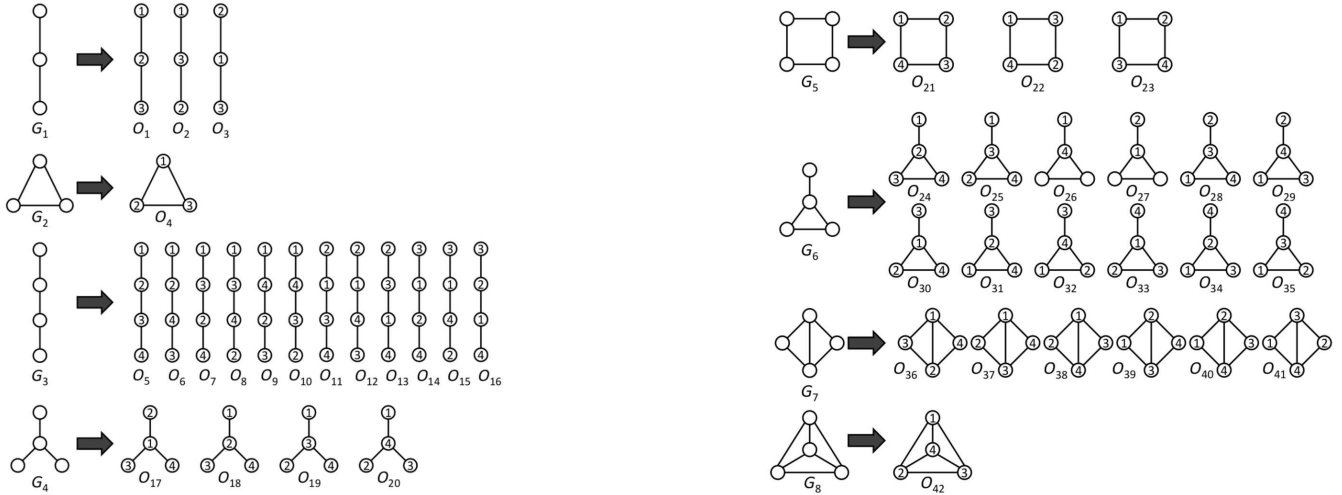}
\caption{All possible eight regular (non-ordered) 3-4-node graphlets ($G_1, G_2, ..., G_8$; on the left of the given arrow) and their corresponding 42 ordered graphlets ($O_1, O_2, ..., O_{42}$; on the right of the given arrow). }
\label{fig:ographlet-all}
\end{figure}

Graphlets are small connected \emph{induced} subgraphs
(Fig. \ref{fig:ographlet-all}). They have been proven as sensitive and
superior measures of topology in numerous contexts when studying
protein-protein interaction networks
\cite{Przulj2004,Przulj2007,Milenkovic2008,Milenkovic2011gdc,Faisal2014dynamic,Malod2014}. Hence,
we use graphlets as PSN measures for PC, as follows.

\vspace{0.1cm}

\noindent\textbf{Graphlet counts.}
We count occurrences of each graphlet on up to $n$ nodes in the given network. To investigate the best choice for $n$, we use counts for 3-4-node (Fig. \ref{fig:ographlet-all}) and 3-5-node graphlets,
resulting in \emph{Graphlet-3-4} and \emph{Graphlet-3-5} measures, respectively. Graphlet counts typically vary by orders of magnitude in real-world networks \cite{Przulj2004}. Hence, we normalize graphlet
counts by taking their logarithms.  Here, we do not consider 3-node-only graphlets, because there are only two 3-node graphlets, which may not be suitable for our PCA framework, and also because using up to 4- or 5-node graphlets improves accuracy upon using only 3-node graphlets \cite{Hulovatyy2014,Hulovatyy2015,Solava2012}.

\vspace{0.1cm}

\noindent\textbf{Normalization of graphlet counts.}
Networks with similar topology can have dissimilar graphlet counts
simply because of their dissimilar network sizes (see Results).  To
remove the bias of PSN size, we normalize graphlet counts by scaling
them between 0 and 1.  Formally, given a network, let $g_1, g_2, ...,
g_n$ be counts of $n$ graphlets $G_1, G_2, ..., G_n$, respectively
($n=8$ for 3-4-node graphlets and $n=29$ for 3-5-node graphlets).  We
normalize count $g_i$ of graphlet $G_i$ as ${g_i}/{\sum_{j=1}^{n}
g_j}$.  We denote the normalized Graphlet-3-4 and Graphlet-3-5
measures as \emph{NormGraphlet-3-4} and \emph{NormGraphlet-3-5},
respectively.
%\vspace{0.1cm}

\noindent\textbf{Integration of graphlets with protein sequences: ordered graphlet counts.}
While amino acids appear in a particular order in the sequence,
graphlets were originally not designed to capture this node order
information. For example, nodes in graphlet $G_1$ can appear in three
different orders (Fig. \ref{fig:ographlet-all}), but $G_1$ cannot
differentiate between them. To take advantage of both network and
sequence data, \emph{ordered graphlets} were recently proposed
\cite{Malod2014}, which embed the \emph{relative} order of nodes onto
graphlets. For example, the three different orders of graphlet $G_1$
were formulated as three different ordered graphlets: $O_1$, $O_2$,
and $O_3$ (Fig. \ref{fig:ographlet-all}).  This way, Malod-Dognin and
Pr\v{z}ulj \cite{Malod2014} defined all four possible 3-node ordered
graphlets for all two possible 3-node ``regular'' (i.e., original
non-ordered) graphlets.

We denote the measure consisting of the existing four counts for
3-node ordered graphlets as \emph{OrderedGraphlet-3}, and we denote
our normalized counterpart of \emph{OrderedGraphlet-3} as
\emph{NormOrderedGraphlet-3} (normalization is done in the same way as 
explained above). Unlike for regular (non-ordered) graphlets, we
\emph{do} consider 3-node-only ordered graphlets within our PCA
approach. We do this to fairly compare our alignment-free PCA approach
with the existing alignment-based non-PCA approach by Malod-Dognin and
Pr\v{z}ulj \cite{Malod2014} that can support only 3-node ordered
graphlets. 
To benefit from larger graphlets, we \emph{extend} this idea to
include within our PCA approach all 38 possible 4-node ordered
graphlets for all six possible 4-node regular graphlets on top of the
existing four 3-node ordered graphlets.  We denote the resulting
measure consisting of 42 ordered graphlet counts for 3-4-node
graphlets (Fig. \ref{fig:ographlet-all}) as
\emph{OrderedGraphlet-3-4} and its normalized counterpart as 
\emph{NormOrderedGraphlet-3-4}.
Inclusion of ordered graphlets on five nodes would cause the number of
graphlets to grow significantly (e.g., graphlet $G_9$ can be
formulated as 60 different ordered graphlets). Since using too many
measures often causes overfitting, which can eventually lead to
increased error rate \cite{Aggarwal2015}, we do not consider 5-node
ordered graphlets. 
Note that ordered graphlet counts do not vary by orders of magnitude
in our data as regular graphlet counts do, so we do not take their
logarithms.

\begin{figure}
  \begin{center} 
  \begin{minipage}[h]{0.1\linewidth}
  \raisebox{-3.5cm}{\epsfig{file=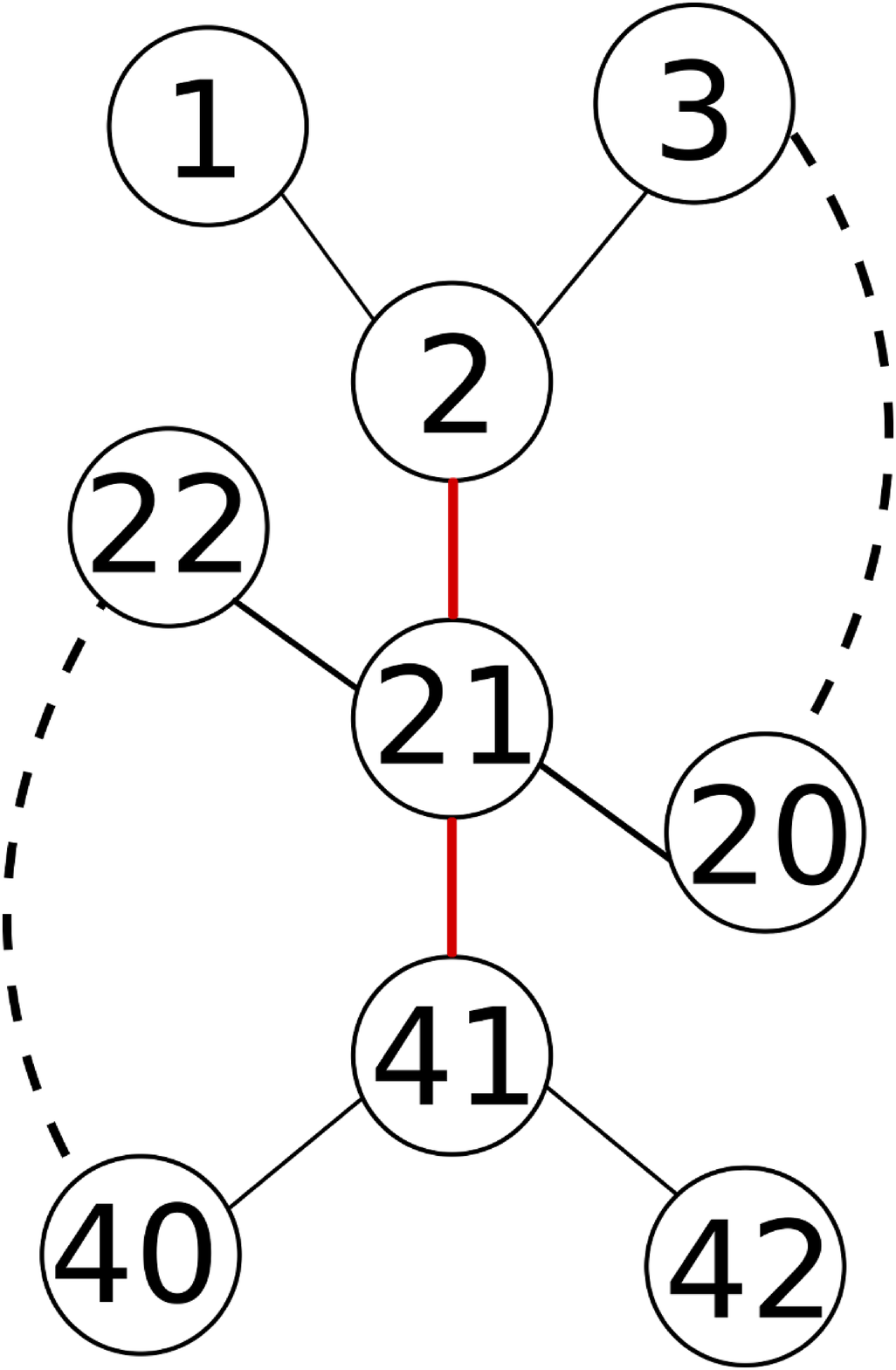,width=\linewidth}} 
  \end{minipage}\hfill
\begin{minipage}[h!]{0.89\linewidth} \caption{
Illustration of the importance of ``long-range($K$)'' ordered
graphlets. A PSN is shown for a toy protein that consists of 42 amino
acids in the sequence, i.e., nodes in the PSN (amino acids 4--19 and
23-39 are not shown for simplicity, as indicated by dashed lines). The
nodes are denoted by their amino acid positions in the sequence. Black
solid lines are network edges that indicate sequence closeness of the
corresponding amino acids (meaning that the amino acids are adjacent
in the sequence), which in turn yields sufficient 3D spacial proximity
of the amino acids. On the other hand, red solid lines are network
edges that indicate only spatial proximity, without sequence
adjacentness. On the one hand, both the three-node path 1--2--3 as
well as the three-node path 2--21--41 correspond to the same ordered
graphlet, namely $O_1$ from Fig. \ref{fig:ographlet-all}, under the
traditional ordered graphlet approach. However, we argue that the
latter is more interesting than the former, as the former is $O_1$
simply because of sequence adjancentness of amino acids 1 and 2 as
well as 2 and 3, while the latter is $O_1$ because of spatial
proximity of amino acids 2 and 21 as well as 21 and 41. On the other
hand, even for $K$ value as low as two, the path 1--2--3 will not be
detected as $O_1$ under the ``long-range($K$)'' ordered graphlet
approach, while the path 2--21--41 will, because all of its linked
node pairs are at least two amino acids apart in the sequence. Note
that the path 2--21--41 will be identified as $O_1$ up to $K$ value of
$min(21-2,41-21)=19$).\label{fig:atleastk} }
\end{minipage} \end{center}
\end{figure}

While ordered graphlets capture relative sequence positions of
interacting amino acids, they do not capture how far those amino acids
are in the sequence.  While there have been conflicting findings regarding 
the effect of long-range interactions on secondary structure prediction accuracy \cite{Kihara2005}, we hypothesize that amino acids that are close
enough in the protein 3D structure but are far away in the protein
sequence are more important than amino acids that are close enough in
the 3D structure simply because they are also close in the sequence.
To capture and evaluate this hypothesis, we propose a novel concept of
\emph{``long-range($K$)'' ordered graphlets}, where the 
``long-range($K$)'' constraint is introduced so that a given ordered
graphlet is identified in the given PSN if and only if: 1) the same
ordered graphlet would also be identified in the above described
analysis, and 2) every pair of amino acids that are linked by an edge
in the graphlet are at least $K$ distance apart in the sequence (that
is, $K$ is the absolute difference between sequence positions of two
amino acids of interest). See Fig. \ref{fig:atleastk} for an
illustration of this concept. Clearly, all graphlets identified under
the ``long-range($K$)'' ordered graphlet approach will also be
identified under the traditional ordered graphlet approach, but the
opposite is not neccesarily true. As a proof of concept, we apply the
concept of ``long-range($K$)'' ordered graphlets on the
\emph{NormOrderedGraphlet-3-4} (which as we will show in Results is the 
best of all graphlet features) and we denote the new measure as
\emph{NormOrderedGraphlet-3-4(K).} 
%XXXXXXXX
To evaluate the performance of
\emph{NormOrderedGraphlet-3-4(K)}, 
we vary $K$ from one to 10 in increments of one and from 10 to 35 in
increments of five. Then, for each considered data set, we report
results for the value of $K$ that results in the best PC accuracy (for
details, see Results).

\subsection{Existing approaches}
\label{method:other}

We use 15 existing \emph{network}, \emph{3D contact}, and
\emph{sequence} approaches in the task of PC (Fig. \ref{fig:dendogram}).

\subsubsection{Existing network approaches}
\label{method:other:network}

Existing approaches of this type that we use for PC (not all of which were proposed for PC but can be adapted to it) can be categorized into graphlet and non-graphlet approaches. None of them use PCA as we do.

\vspace{0.1cm}

\noindent\textbf{Existing graphlet  approaches.}
These include graphlet degree distribution agreement (GDDA)
\cite{Przulj2007}, relative graphlet frequency distance (RGFD) \cite{Przulj2004},  graphlet correlation distance (GCD) \cite{Yaveroglu2014}, and GR-align \cite{Malod2014}.  Among them,  GDDA, RGFD, and GCD can compare any type of networks, while GR-align has been specifically designed to compare PSNs. GDDA, RGFD, and GCD
are alignment-free network comparison approaches, while GR-align is an alignment-based approach. In particular, it was the GR-align study \cite{Malod2014} that introduced the idea of 3-node-only ordered graphlets, which we partly base our approach on.

Two alternative graphlet approaches were used in the context of PSNs
\cite{Vacic2010,lugomartinez2014}, but they were used to predict
(classify in a supervised manner) functional residues in PSNs (where
residues are nodes in PSNs) and not for PSN comparison. Since these
approaches compare nodes rather than networks, and since they are
supervised (while our study is unsupervised, per our discussion
below), the approaches do not fit the context of our study. As such,
we do not consider them further.

\vspace{0.1cm}

\noindent\textbf{Existing non-graphlet approaches.}
Several PSN measures have already been used for PC:
\emph{average degree}, \emph{average distance}, \emph{maximum distance}, \emph{average closeness centrality}, \emph{average clustering coefficient},
\emph{intra-hub connectivity}, and \emph{assortativity} (Supplementary Section S1.2)%\ref{supple:method:existing-features})
\cite{Pabuwal2008,Pabuwal2009,Gao2009,Emerson2012,Emerson2013}. For each measure, for each pair of networks, we compute Euclidean distance between the networks' vectors (because all vectors are 1-dimensional, here we cannot use cosine similarity as for our approach).

We combine the seven measures into an eighth measure,
\emph{Existing-all}, to investigate whether the integration of different and complementary topological measures helps PC. We use Existing-all within our PCA framework. This way, we can fairly compare our graphlet measures and the existing non-graphlet measures within the same framework.

\subsubsection{Existing 3D contact approaches}
\label{method:other:3dcontact}

These include DaliLite \cite{DaliLite2010} and TM-align
\cite{TMAlign2005}, both of which are alignment-based.

\subsubsection{Existing sequence approaches}
\label{method:other:sequence}

Mizianty et al. \cite{Mizianty2010} used protein length and amino acid
propensities to define a sequence measure, which outperformed
methodologies of 11 other methodologies \cite{Mizianty2010}.  Thus, we
use this measure, denoted as \emph{Sequence}, within our PCA
framework.  This way, we can fairly compare network and sequence
measures within the same framework.

\subsection{Evaluation of PC accuracy}
\label{method:evaluation}

Given a set of objects (proteins or networks) with known labels, for a
good approach, the distance between objects of the same label should
be small, while the distance between objects of different labels
should be large.  To evaluate this, we rely on an established
unsupervised strategy \cite{Yaveroglu2015}. By ``unsupervised'', we
mean that we rely on object labels only in the phase of evaluating a
method's output. That is, we do not use any label information to train
the given method or produce its output, as a supervised
(classification) approach would do. Details of our evaluation are as
follows.  For each approach, we first compute the distance between
each pair of objects according to the given PC approach.  Then, we
sort all object pairs in terms of their increasing distance and
consider $k$ closest object pairs, where we vary $k$ from 0\% to 100\%
in increments of 0.1\%.  Next, we compute the accuracy in terms of
\emph{precision} and
\emph{recall}, where precision is the fraction of label-matching
object pairs out of the considered object pairs, and recall is the
fraction of the considered label-matching object pairs out of all
label-matching object pairs.  To summarize the precision and recall
results over the whole [0-100\%] range of $k$, we measure overall
accuracy of the given PC approach by computing AUPR.  Alternatively,
we compute the accuracy in terms of \emph{sensitivity} and
\emph{specificity}, where sensitivity is the fraction of the
considered label-matching object pairs out of all label-matching
object pairs, and specificity is the fraction of the considered
non-label-matching object pairs out of all non-label-matching object
pairs.  To summarize the sensitivity and specificity results over the
whole [0-100\%] range of $k$, we measure overall accuracy of the given
approach by computing AUROC.  Given a data set, we compare different
approaches by comparing their AUPR or AUROC scores.

\section{Results}
\label{results}

\subsection{Comparison of synthetic networks}
\label{results:synthetic}

The motivation behind evaluating our approach against the existing ones on synthetic networks is to demonstrate the general applicability of our approach to any domain where data can be modeled with networks. This is important, because some of the existing approaches that we evaluate against have already been used in tasks different than our considered task of PC. So, if we can demonstrate the
superiority of our approach over such widely applicable existing approaches, then this would imply that henceforth it is our approach that should be used in those tasks instead.

Unlike for real-world PSNs, for synthetic networks, we cannot evaluate 3D contact and sequence approaches, as they require 3D contact and sequence information, respectively, which synthetic networks do not
contain. Thus, we can only apply network approaches to synthetic networks, with the exception of ordered graphlet approaches, including GR-align, that require some node order, which again synthetic networks do not have. We evaluate the remaining (15) network approaches on synthetic networks. For these networks, the topology-based ground truth label categorization is known. (Note that a random network model that would, unlike the existing general-purpose random network models that we use, generate synthetic networks with embedded node order which would closely mimic all of protein sequence, 3D, and network structure, would fit better the context of PSN comparison than the general-purpose models that we use. However, developing such a model is non-trivial and is thus out of the scope of the current study.)

First, we evaluate the network approaches (i.e., their existing
versions that are non-normalized in terms of network size) on
synthetic networks of the same size.  Second, we evaluate whether the
current non-normalized versions of the network approaches can
successfully cope with synthetic networks of different sizes.  We find
that our graphlet PCA approach overall outperforms the existing
network approaches, including the existing graphlet (non-PCA)
approaches. Therefore, in all subsequent tests, we focus on the
graphlet PCA methodology.  Yet, the accuracy of the graphlet PCA
approach (as well as every other approach) drops when analyzing
networks of different sizes compared to analyzing networks of the same
size, meaning that some level of miscategorization arises due to the
networks having different sizes. This indicates a need for devising a
normalized version of the graphlet PCA approach.  Thus, third, we
develop such a normalized approach, and as we show, normalization
indeed improves PC.  Forth, we summarize our key findings resulting
from analyzing the synthetic network data. The four items are
discussed in the following four subsections.

\subsubsection{Evaluation of  non-normalized network measures}
\label{results:synthetic:benchmarking}

Here, we evaluate non-normalized versions of our graphlet PCA approach
(i.e., Graphlet-3-4 and Graphlet-3-5), existing graphlet approaches
(i.e., GDDA, RGFD, and GCD), and existing non-graphlet approaches
(i.e., average degree, average distance, maximum distance, average
closeness centrality, average clustering coefficient, intra-hub
connectivity, assortativity, and Existing-all).  We evaluate the
approaches on synthetic network data of the same size (i.e.,
Synthetic-100, Synthetic-500, and Synthetic-1000). For details, see
Methods.

For each data set, both non-normalized versions of our graphlet PCA
approach outperform the existing graphlet and non-graphlet approaches,
as the former two achieve 100\% accuracy (Table
\ref{tab:synthetic-aupr} and Supplementary Table S2).
%\ref{supple:tab:synthetic-auroc}). 
Some of the existing methods also achieve 100\% accuracy on some of the data sets,
but only one (RGFD) does so on all three data sets and is thus
comparable to our approach. However, as we show below, RGFD loses
its comparable performance in other tests.

Both our graphlet PCA approach and the existing graphlet (non-PCA) approaches outperform the existing non-graphlet approaches (Table \ref{tab:synthetic-aupr} and Supplementary Table S2).
%\ref{supple:tab:synthetic-auroc}).  
This confirms the power of the local graphlets over the global network measures that have traditionally been used for PC.

Combining the seven existing non-graphlet measures into Existing-all
and using Existing-all in our PCA framework improves the accuracy of
each individual non-graphlet measure.  This confirms that measure
integration helps, which is why we have developed our framework to
allow for this in the first place.  Existing-all is comparable to our
graphlet PCA approach (Table \ref{tab:synthetic-aupr} and
Supplementary Table S2).
%\ref{supple:tab:synthetic-auroc}). 
However, Existing-all loses its
comparable performance in other tests (see below).

\subsubsection{Network size affects comparison via non-normalized measures}
\label{results:synthetic:network-bias}

To test whether the non-normalized versions of our graphlet PCA
approach, existing graphlet approaches, and existing non-graphlet
approaches, all of which are non-normalized, are robust to the size of
networks to be compared, we evaluate the approaches on the
Synthetic-all set, which contains networks with different topologies
\emph{and} of different sizes (unlike the equal-size network sets from the above analysis).
In this analysis, we observe a decline in accuracy for each approach
(Table
\ref{tab:synthetic-aupr} and Supplementary Table S2).
%\ref{supple:tab:synthetic-auroc}). 
Clearly, the accuracy is biased by network size.

\subsubsection{Normalization of graphlet measures improves comparison}
\label{results:synthetic:normalization}

Motivated by this network size-related bias of all non-normalized
network measures, we propose a normalized version of the best of all
measures, namely our graphlet PCA measures.  We validate our
normalized graphlet PCA measures as follows. When we apply them to
Synthetic-100, Synthetic-500, and Synthetic-1000, we hope to preserve
the maximum (100\%) accuracy for the three network data sets of the
same size while improving the accuracy for Synthetic-all that contains
networks of different sizes. Indeed, this is exactly what we observe
(Table \ref{tab:synthetic-aupr} and Supplementary Table S2).
%\ref{supple:tab:synthetic-auroc}).  
Now the best of our graphlet PCA approaches (i.e., NormGraphlet-3-5)
outperforms each of the three existing graphlet (non-PCA) approaches,
even though all of these approaches are based on graphlets. This shows
the usefulness of our PCA framework as a whole over the existing
graphlet methodologies. Also, now NormGraphlet-3-5 outperforms the
non-graphlet Existing-all approach under the same PCA framework, which
confirms the power of graphlets.

\begin{SCtable}
%\begin{center} 
\caption{Accuracy with respect to AUPRs 
(expressed as percentages) on synthetic networks. Results for
non-normalized approaches are highlighted in 1) light gray for network
data of the same size and 2) dark gray for network data of different
sizes.  Results for normalized approaches are not highlighted.  Given
a network data set (within a column), the AUPR of the best approach is
shown in bold.  For equivalent results with respect to AUROCs, see
Supplementary Table S2.
%\ref{supple:tab:synthetic-auroc}.
} 
\fontsize{7}{9}\selectfont
\begin{tabular}{|l|cccc|}
\hline
{} & \multicolumn{4}{c|}{Synthetic}\\
\cline{2-5}
Approach & Synthetic- & Synthetic- & Synthetic- & Synthetic- \\
{} & 100 & 500 & 1000 & All \\
\hline
Graphlet-3-4 & \cellcolor{lightgray} \textbf{100.00} & \cellcolor{lightgray} \textbf{100.00} & \cellcolor{lightgray} \textbf{100.00} & \cellcolor{gray} 81.76 \\
Graphlet-3-5 & \cellcolor{lightgray} \textbf{100.00} & \cellcolor{lightgray} \textbf{100.00} & \cellcolor{lightgray} \textbf{100.00} & \cellcolor{gray} 83.28 \\
NormGraphlet-3-4 & \textbf{100.00} & \textbf{100.00} & \textbf{100.00} & 94.37 \\
NormGraphlet-3-5 & \textbf{100.00} & \textbf{100.00} & \textbf{100.00} & \textbf{99.86} \\
\hline
GDDA & \cellcolor{lightgray} 97.36 & \cellcolor{lightgray} \textbf{100.00} & \cellcolor{lightgray} 99.99 & \cellcolor{gray} 91.46 \\
RGFD & \cellcolor{lightgray} \textbf{100.00} & \cellcolor{lightgray} \textbf{100.00} & \cellcolor{lightgray} \textbf{100.00} & \cellcolor{gray} 98.55 \\
GCD & \cellcolor{lightgray} 89.26 & \cellcolor{lightgray} \textbf{100.00} & \cellcolor{lightgray} \textbf{100.00} & \cellcolor{gray} 86.27 \\
\hline
Average degree & \cellcolor{lightgray} 79.76 & \cellcolor{lightgray} 79.76 & \cellcolor{lightgray} 79.76 & \cellcolor{gray} 68.77 \\
Average distance & \cellcolor{lightgray} 82.47 & \cellcolor{lightgray} 98.12 & \cellcolor{lightgray} 99.60 & \cellcolor{gray} 57.10 \\
Maximum distance & \cellcolor{lightgray} 68.82 & \cellcolor{lightgray} 84.32 & \cellcolor{lightgray} 93.08 & \cellcolor{gray} 46.11 \\
Average closeness centrality & \cellcolor{lightgray} 86.10 & \cellcolor{lightgray} 88.46 & \cellcolor{lightgray} 85.33 & \cellcolor{gray} 48.41 \\
Average clustering coefficient & \cellcolor{lightgray} 98.93 & \cellcolor{lightgray} 99.68 & \cellcolor{lightgray} 99.25 & \cellcolor{gray} 79.37 \\
Intra-hub connectivity & \cellcolor{lightgray} 70.88 & \cellcolor{lightgray} 69.11 & \cellcolor{lightgray} 69.31 & \cellcolor{gray} 66.61 \\
Assortativity & \cellcolor{lightgray} 82.79 & \cellcolor{lightgray} 92.27 & \cellcolor{lightgray} 91.73 & \cellcolor{gray} 81.98 \\
Existing-all & \cellcolor{lightgray} \textbf{100.00} & \cellcolor{lightgray} \textbf{100.00} & \cellcolor{lightgray} \textbf{100.00} & \cellcolor{gray} 85.92 \\
\hline
\end{tabular}
\label{tab:synthetic-aupr}
%\end{center} 
\end{SCtable}

\subsubsection{Summary of results for synthetic networks}
\label{results:synthetic:graphlet-vs-nongraphlet}

Our (non-normalized) graphlet PCA measures overall outperform the
existing graphlet (non-PCA) approaches, which in turn outperform the
existing non-graphlet approaches.  Our normalized graphlet PCA
measures further improve upon their non-normalized counterparts (and
thus upon the existing approaches).  NormGraphlet-3-5 is the most
accurate approach.

\subsection{Comparison of PSNs}
\label{results:real}

In our analysis of real-world PSNs, for which CATH- or
SCOP-label-based (rather than topology-based as above) ground truth
label categorization is known, first, we evaluate the approaches
(i.e., their existing versions that are non-normalized in terms of
network size) on PSNs of the same size.  Second, we test the
approaches on PSNs of different sizes.  In both tests, overall, our
graphlet PCA approach is superior to the existing approaches. Yet, the
accuracy of all approaches drops when analyzing PSNs of different
sizes compared to analyzing PSNs of the same size.  Therefore, third,
we test whether graphlet normalization improves PC.  Indeed, this is
what we observe.  Fourth, to investigate whether the integration of
network topology with protein sequences can improve PC, we test our
ordered graphlet PCA approach, including the effect of the
``long-range($K$)'' constraint. Fifth, we compare the considered
approaches in terms of their running times. Sixth, we summarize our
key findings resulting from analyzing the PSN data. The six items are
discussed in the following six subsections.

\subsubsection{Evaluation of  non-normalized measures}
\label{results:real:benchmarking}

Here, we benchmark the non-normalized versions of our PCA graphlet
approach, existing graphlet (non-PCA) approaches, existing
non-graphlet approaches, and 3D contact approaches on all PSN data
sets for which networks within the given set are of the same size,
i.e., on CATH-95, CATH-99, and CATH-251-265.  For each PSN set, just
as for the synthetic networks, the non-normalized versions of our
graphlet PCA approach (Graphlet-3-4 and Graphlet-3-5) are superior to
the existing graphlet, non-graphlet, and 3D contact approaches, except
one (RGFD) that is comparable to our approach (Table
\ref{tab:real-aupr} and Supplementary Table S3).
%\ref{supple:tab:real-auroc}). 
Yet, as we show below, RGFD loses its comparable performance in other
tests. Again, combining the seven existing non-graphlet measures into
Existing-all typically improves the accuracy of each individual
measure (Table \ref{tab:real-aupr}). Existing-all is comparable to the two non-normalized versions of our graphlet PCA approach. However, as we show below, Existing-all loses its comparable performance in other tests.

\subsubsection{Network size affects comparison via non-normalized measures} 
\label{results:real:network-bias}

Next, we evaluate the same non-normalized approaches on all 10 sets of
PSNs of different sizes (i.e., CATH-primary, CATH-$\alpha$,
CATH-$\beta$, CATH-$\alpha$/$\beta$, SCOP-primary, SCOP-$\alpha$,
SCOP-$\beta$, SCOP-$\alpha$/$\beta$, SCOP-$\alpha$+$\beta$, and
SCOP-multidomain).  We observe a decline in accuracy for each
approach, which confirms the bias of network size. Nonetheless, the
non-normalized versions of our graphlet PCA approach remain superior
or comparable to all existing methods (Table \ref{tab:real-aupr} and
Supplementary Table S3).
%\ref{supple:tab:real-auroc}).

\subsubsection{Normalization of graphlet  measures improves comparison}
\label{results:real:normalization}

Motivated by the above network size-related bias of all considered
non-normalized approaches, we propose a normalized version of the best
of those approaches, namely of the graphlet PCA measures. When we
apply each of the normalized measures to the PSN data, we hope to
ideally improve or at least preserve the accuracy on the PSN data sets
of the same network size (i.e., CATH-95, CATH-99, and CATH-251-265)
while improving the accuracy for the 10 sets of PSNs of different
sizes, compared to the accuracy of the measures' non-normalized
counterparts.  Indeed, this is exactly what we observe (Table
\ref{tab:real-aupr} and Supplementary Table S3).
%\ref{supple:tab:real-auroc}).

\begin{table}[]
\begin{center} 
\caption{Accuracy with respect to AUPRs (expressed as 
percentages) on real-world PSN data sets of the same network size as
well as of different network sizes. Results for non-normalized
approaches are highlighted in 1) light gray for network data of the
same size and 2) dark gray for network data of different sizes.
Results for normalized approaches are not highlighted. Given a network
data set (within a given column), the AUPR of the best approach is
shown in bold.  For equivalent results with respect to AUROCs, see
Supplementary Table S3.
%\ref{supple:tab:real-auroc}.
}
\fontsize{7}{9}\selectfont
\begin{tabular}{|l|ccc|cccc|cccccc|}
\hline
{} & \multicolumn{3}{c|}{CATH of the same size} &
\multicolumn{4}{c|}{CATH (of different sizes)} & \multicolumn{6}{c|}{SCOP (of different sizes)} \\
\cline{2-14}
Approach & CATH- & CATH- & CATH- & primary & $\alpha$ & $\beta$ & $\alpha$/$\beta$ & primary & $\alpha$ & $\beta$ & $\alpha$/$\beta$ & $\alpha$+$\beta$ & Multi \\
{} & 95 & 99 & 251-265 & {} & {} & {} & {} & {} & {} & {} & {} & {} & domain \\
\hline
Graphlet-3-4 & \cellcolor{lightgray} 82.28 & \cellcolor{lightgray} 92.05 & \cellcolor{lightgray} 92.35 & \cellcolor{gray} 47.47 & \cellcolor{gray} 50.53 & \cellcolor{gray} 38.50 & \cellcolor{gray} 46.62 & \cellcolor{gray} 37.85 & \cellcolor{gray} 19.95 & \cellcolor{gray} 32.59 & \cellcolor{gray} 21.68 & \cellcolor{gray} 15.37 & \cellcolor{gray} 63.46 \\
Graphlet-3-5 & \cellcolor{lightgray} 83.31 & \cellcolor{lightgray} 92.78 & \cellcolor{lightgray} 92.89 & \cellcolor{gray} 47.86 & \cellcolor{gray} 50.46 & \cellcolor{gray} 37.78 & \cellcolor{gray} 45.77 & \cellcolor{gray} 37.29 & \cellcolor{gray} 19.57 & \cellcolor{gray} 26.47 & \cellcolor{gray} 17.99 & \cellcolor{gray} 12.77 & \cellcolor{gray} 63.86 \\
OrderedGraphlet-3 & \cellcolor{lightgray} 90.99 & \cellcolor{lightgray} 95.93 & \cellcolor{lightgray} 91.02 & \cellcolor{gray} 45.75 & \cellcolor{gray} 52.19 & \cellcolor{gray} 42.04 & \cellcolor{gray} \textbf{49.69} & \cellcolor{gray} 35.33 & \cellcolor{gray} 21.71 & \cellcolor{gray} 44.71 & \cellcolor{gray} 19.80 & \cellcolor{gray} 20.24 & \cellcolor{gray} 60.37 \\
OrderedGraphlet-3-4 & \cellcolor{lightgray} 96.69 & \cellcolor{lightgray} 91.88 & \cellcolor{lightgray} 97.20 & \cellcolor{gray} \cellcolor{gray} 46.16 & \cellcolor{gray} 52.39 & \cellcolor{gray} 40.50 & \cellcolor{gray} 48.38 & \cellcolor{gray} 33.60 & \cellcolor{gray} 20.72 & \cellcolor{gray} 39.78 & \cellcolor{gray} 17.37 & \cellcolor{gray} 16.29 & \cellcolor{gray} 60.19 \\
NormGraphlet-3-4 & 96.03 & \textbf{100.00} & 95.28 & 52.57 & 50.93 & 37.59 & 43.24 & 36.89 & 17.15 & 23.75 & 18.59 & 13.66 & 68.95 \\
NormGraphlet-3-5 & 94.11 & 99.73 & 97.67 & 53.24 & 51.02 & 37.21 & 43.74 & 37.51 & 17.31 & 23.55 & 18.03 & 9.85 & 67.51 \\
NormOrderedGraphlet-3  & 83.54 & 96.49 & 93.51 & 54.2 & 52.79 & 47.24 & 33.5 & 33.36 & 15.26 & 18.72 & 12.95 & 9.39 & 67.23 \\
NormOrderedGraphlet-3-4 & \textbf{97.59} & 96.58 & \textbf{98.74} & \textbf{65.64} & 53.05 & 49.11 & 44.34 & \textbf{44.50} & 21.75 & 26.23 & 16.68 & 19.92 & 72.70 \\
NormOrderedGraphlet-3-4(K) & \textbf{97.59} & 96.58 & \textbf{98.74} & \textbf{65.64} & \textbf{53.41} & 49.11 & 48.31 & \textbf{44.50} & \textbf{31.63} & 38.83 & \textbf{34.35} & 33.84 & \textbf{82.02} \\
\hline
GDDA & \cellcolor{lightgray} 77.65 & \cellcolor{lightgray} 80.78 & \cellcolor{lightgray} 71.46 & \cellcolor{gray} 42.77 & \cellcolor{gray} 49.09 & \cellcolor{gray} 34.36 & \cellcolor{gray} 34.15 & \cellcolor{gray} 32.09 & \cellcolor{gray} 14.77 & \cellcolor{gray} 9.99 & \cellcolor{gray} 11.92 & \cellcolor{gray} 5.77 & \cellcolor{gray} 66.79 \\
RGFD & \cellcolor{lightgray} 87.87 & \cellcolor{lightgray} 89.49 & \cellcolor{lightgray} 94.00 & \cellcolor{gray} 53.75 & \cellcolor{gray} 51.26 & \cellcolor{gray} 42.15 & \cellcolor{gray} 41.77 & \cellcolor{gray} 38.86 & \cellcolor{gray} 17.42 & \cellcolor{gray} 21.15 & \cellcolor{gray} 12.90 & \cellcolor{gray} 9.76 & \cellcolor{gray} 66.21 \\
GCD & \cellcolor{lightgray} 71.70 & \cellcolor{lightgray} 74.92 & \cellcolor{lightgray} 77.23 & \cellcolor{gray} 42.61 & \cellcolor{gray} 49.99 & \cellcolor{gray} 37.00 & \cellcolor{gray} 31.26 & \cellcolor{gray} 31.78 & \cellcolor{gray} 14.20 & \cellcolor{gray} 11.46 & \cellcolor{gray} 10.62 & \cellcolor{gray} 6.37 & \cellcolor{gray} 68.33 \\
GR-align & \cellcolor{lightgray} 76.25 & \cellcolor{lightgray} 65.03 & \cellcolor{lightgray} 70.25 & \cellcolor{gray} 39.07 & \cellcolor{gray} 52.68 & \cellcolor{gray} \textbf{50.11} & \cellcolor{gray} 45.03 & \cellcolor{gray} 28.55 & \cellcolor{gray} 30.05 & \cellcolor{gray} \textbf{59.07} & \cellcolor{gray} 24.17 & \cellcolor{gray} \textbf{41.17} & \cellcolor{gray} 77.43 \\
\hline
Average degree & \cellcolor{lightgray} 48.22 & \cellcolor{lightgray} 50.21 & \cellcolor{lightgray} 61.22 & \cellcolor{gray} 42.81 & \cellcolor{gray} 50.22 & \cellcolor{gray} 37.16 & \cellcolor{gray} 38.55 & \cellcolor{gray} 30.85 & \cellcolor{gray} 11.54 & \cellcolor{gray} 14.31 & \cellcolor{gray} 11.12 & \cellcolor{gray} 5.98 & \cellcolor{gray} 60.37 \\
Average distance & \cellcolor{lightgray} 64.49 & \cellcolor{lightgray} 60.22 & \cellcolor{lightgray} 51.60 & \cellcolor{gray} 35.13 & \cellcolor{gray} 50.99 & \cellcolor{gray} 38.61 & \cellcolor{gray} 34.78 & \cellcolor{gray} 29.75 & \cellcolor{gray} 12.51 & \cellcolor{gray} 24.76 & \cellcolor{gray} 14.06 & \cellcolor{gray} 8.21 & \cellcolor{gray} 72.98 \\
Maximum distance & \cellcolor{lightgray} 62.39 & \cellcolor{lightgray} 73.49 & \cellcolor{lightgray} 54.89 & \cellcolor{gray} 35.07 & \cellcolor{gray} 49.59 & \cellcolor{gray} 37.35 & \cellcolor{gray} 39.78 & \cellcolor{gray} 29.67 & \cellcolor{gray} 15.50 & \cellcolor{gray} 19.21 & \cellcolor{gray} 13.95 & \cellcolor{gray} 8.37 & \cellcolor{gray} 58.33 \\
Average closeness centrality & \cellcolor{lightgray} 62.73 & \cellcolor{lightgray} 60.94 & \cellcolor{lightgray} 45.73 & \cellcolor{gray} 34.67 & \cellcolor{gray} 49.86 & \cellcolor{gray} 36.60 & \cellcolor{gray} 39.55 & \cellcolor{gray} 27.84 & \cellcolor{gray} 12.23 & \cellcolor{gray} 21.32 & \cellcolor{gray} 11.58 & \cellcolor{gray} 7.55 & \cellcolor{gray} 70.54 \\
Average clustering coefficient & \cellcolor{lightgray} 87.01 & \cellcolor{lightgray} 72.10 & \cellcolor{lightgray} 89.96 & \cellcolor{gray} 44.15 & \cellcolor{gray} 49.19 & \cellcolor{gray} 39.17 & \cellcolor{gray} 32.28 & \cellcolor{gray} 30.77 & \cellcolor{gray} 11.47 & \cellcolor{gray} 16.26 & \cellcolor{gray} 10.58 & \cellcolor{gray} 6.61 & \cellcolor{gray} 63.47 \\
Intra-hub connectivity & \cellcolor{lightgray} 54.94 & \cellcolor{lightgray} 72.34 & \cellcolor{lightgray} 63.76 & \cellcolor{gray} 34.79 & \cellcolor{gray} 50.14 & \cellcolor{gray} 36.85 & \cellcolor{gray} 40.04 & \cellcolor{gray} 29.21 & \cellcolor{gray} 11.87 & \cellcolor{gray} 23.41 & \cellcolor{gray} 14.26 & \cellcolor{gray} 10.66 & \cellcolor{gray} 64.51 \\
Assortativity & \cellcolor{lightgray} 76.97 & \cellcolor{lightgray} 85.34 & \cellcolor{lightgray} 93.31 & \cellcolor{gray} 37.41 & \cellcolor{gray} 48.10 & \cellcolor{gray} 35.90 & \cellcolor{gray} 30.64 & \cellcolor{gray} 26.99 & \cellcolor{gray} 8.98 & \cellcolor{gray} 12.11 & \cellcolor{gray} 9.02 & \cellcolor{gray} 5.18 & \cellcolor{gray} 54.93 \\
Existing-all & \cellcolor{lightgray} 82.14 & \cellcolor{lightgray} 91.56 & \cellcolor{lightgray} 92.48 & \cellcolor{gray} 47.69 & \cellcolor{gray} 50.46 & \cellcolor{gray} 36.53 & \cellcolor{gray} 41.35 & \cellcolor{gray} 35.68 & \cellcolor{gray} 19.49 & \cellcolor{gray} 23.56 & \cellcolor{gray} 14.16 & \cellcolor{gray} 9.33 & \cellcolor{gray} 76.35 \\
\hline
DaliLite & \cellcolor{lightgray} 53.38 & \cellcolor{lightgray} 69.12 & \cellcolor{lightgray} 58.96 & \cellcolor{gray} 39.84 & \cellcolor{gray} 49.22 & \cellcolor{gray} 46.34 & \cellcolor{gray} 39.33 & \cellcolor{gray} 31.64 & \cellcolor{gray} 18.63 & \cellcolor{gray} 27.85 & \cellcolor{gray} 22.16 & \cellcolor{gray} 13.04 & \cellcolor{gray} 70.22 \\
TM-align & \cellcolor{lightgray} 50.93 & \cellcolor{lightgray} 62.02 & \cellcolor{lightgray} 45.79 & \cellcolor{gray} 37.30 & \cellcolor{gray} 45.56 & \cellcolor{gray} 34.77 & \cellcolor{gray} 28.63 & \cellcolor{gray} 25.78 & \cellcolor{gray} 8.05 & \cellcolor{gray} 16.99 & \cellcolor{gray} 11.45 & \cellcolor{gray} 7.55 & \cellcolor{gray} 58.18 \\
\hline
Sequence & \cellcolor{lightgray} 70.23 & \cellcolor{lightgray} 62.14 & \cellcolor{lightgray} 54.48 & \cellcolor{gray} 40.45 & \cellcolor{gray} 50.24 & \cellcolor{gray} 37.66 & \cellcolor{gray} 35.18 & \cellcolor{gray} 29.02 & \cellcolor{gray} 22.13 & \cellcolor{gray} 20.74 & \cellcolor{gray} 11.82 & \cellcolor{gray} 11.77 & \cellcolor{gray} 73.78 \\
\hline
\end{tabular}
\label{tab:real-aupr} 
\end{center}
\end{table}

\subsubsection{Integration of network and sequence data via ordered graphlets}
\label{results:real:ordered-graphlet}

The versions of our PCA approach that are based on regular
(non-ordered, as considered thus far) graphlets, already perform much
better than the sequence approach (Table
\ref{tab:real-aupr} and Supplementary Table S3).
%\ref{supple:tab:real-auroc}). 
Integration of network data with sequence data may further improve the accuracy compared to only network and only sequence approaches.  We test this by using
ordered graphlets to impose the sequence-based order of amino acids
onto nodes in regular graphlets (Fig. \ref{fig:ographlet-all}).

Considering only non-normalized graphlet measures, ordered graphlets
(i.e., OrderedGraphlet-3 and OrderedGraphlet-3-4) improve upon their
regular graphlet counterparts for PSNs of the same size as well as of
different sizes (Table \ref{tab:real-aupr}).  Considering also normalized graphlet measures,
NormOrderedGraphlet-3-4 leads to better accuracy compared to its
non-normalized counterpart, though NormOrderedGraphlet-3 does not
improve upon its non-normalized counterpart (Table
\ref{tab:real-aupr}).

Integrating the ``long-range($K$)'' constraint on top of
\emph{NormOrderedGraphlet-3-4}, i.e., considering
\emph{NormOrderedGraphlet-3-4(K)}, further improves 
accuracy (Table \ref{tab:real-aupr} and Supplementary Table S3).
%\ref{supple:tab:real-auroc}). 
Recall that in these tests, we vary $K$ 
(see Methods). The best value of $K$ is data set-dependent. Of the 13
considered data sets, increasing $K$ to at least two (i.e., considering
the ``long-range($K$)'' ordered graphlet approach) helps compared to
$K=1$ (i.e., compared to the traditional ordered graphlet approach)
for the majority (seven) of the data sets (Supplementary Tables S4 and S5).
%\ref{supple:tab:atleast-aupr} and \ref{supple:tab:atleast-auroc}). 
In particular, there is significant increase in accuracy for all data
sets corresponding to the secondary hierarchical categories of CATH
(except \emph{CATH-$\beta$}) and SCOP. For the seven data sets, the
best value of $K$ ranges from three to 35. Since even as high value of $K$
as 35 yields better accuracy than smaller values of $K$, these results
exemplify the importance of long-range interactions in the task of PC.

The fact that within our PCA framework ordered graphlets beat regular
graphlets alone and the sequence approach alone confirms that PSN data
and sequence data are complementary and should thus be integrated. We
consider this to be one of our key contributions.  Here, we note that
3-node-only ordered graphlets
\emph{were} used for protein 3D structural alignment within the
GR-align approach. We adopt the existing idea of 3-node ordered
graphlets but we do so within our alignment-free PCA framework as
opposed to the existing alignment-based GR-align approach. Also, we
extend this idea into larger, 3-4-node ordered graphlets. Further, we
add a ``long-range($K$)'' constraint into the process of ordered
graphlet counting. Importantly, when we consider 3-node-only ordered
graphlets within our PCA framework, which makes the comparison with
GR-align as fair as possible, our PCA approach is superior to GR-align
(Table \ref{tab:real-aupr}). This is further supported when we compare the accuracy
rankings of the different methods over all PSN sets (Table
\ref{tab:running-time}). When we also consider larger ordered
graphlets, this further improves the performance of our PCA approach,
and so does the ``long-range($K$)'' ordered graphlet constraint (Tables
\ref{tab:real-aupr} and \ref{tab:running-time}). 
GR-align is also slower than our approach. For example, it is 25 times
slower than the fairly comparable 3-node-only ordered graphlet version
of our PCA approach (Table \ref{tab:running-time}).  These results
validate our PCA framework as a whole.  Note that when one's goal is
not just to quantify the level of similarity between networks but also
to map nodes between the networks, using an alignment-free approach
such as our graphlet PCA framework is inappropriate, and instead, an
alignment-based approach such as GR-align needs to be used. For
details on alignment-free versus alignment-based approaches, see
Yaveroglu \emph{et al.} \cite{Yaveroglu2015}.

\begin{SCtable}
%\begin{center} 
\caption
{Summary of method accuracy and running times. Accuracy of the given
approach is shown with respect to its average ranking compared to all
considered approaches across all considered real-world PSN sets, and
the results are shown based on AUPR as well as AUROC.  The ranking of
each method is expressed as follows. For the given PSN set, we
determine which approach results in the highest accuracy (rank 1), the
second highest accuracy (rank 2), etc. Then, we average the rankings
of the given method over all PSN sets. So, the lower the average rank,
the better the method. Since NormOrderedGraphlet-3-4(K) has the best
average rank with respect to both AUPR and AUROC (shown in bold), we
compute the statistical significance of the improvement of
NormOrderedGraphlet-3-4(K) over each of the other approaches in terms
of their ranks, using paired $t$-test.  Running times of the
approaches are shown when comparing proteins from the CATH-$\alpha$
set. Running times for the other data sets are qualitatively the same.
For visual representation of the results, see Supplementary
Fig. S1 and S2.
%\ref{supple:fig:rank-aupr} and \ref{supple:fig:rank-auroc}.
}
\fontsize{7}{9}\selectfont
\begin{tabular}{|l|rr|rr|r|}
\hline
Approach & \multicolumn{2}{c|}{AUPR} & \multicolumn{2}{c|}{AUROC} &
Running \\ {} & Rank & $p$-value & Rank & $p$-value & time (hrs) \\
\hline
Graphlet-3-4 & 8.38 & 9.42e-05 & 10.50 & 0.000147 & 0.43 \\ 
  Graphlet-3-5 & 9.00 & 4.81e-06 & 10.40 & 8.74e-05 & 0.49 \\ 
  OrderedGraphlet-3 & 7.15 & 0.00225 & 9.92 & 0.000692 & 0.38\\ 
  OrderedGraphlet-3-4 & 7.31 & 0.00143 & 8.69 & 0.0018 & 2.39\\ 
  NormGraphlet-3-4 & 7.77 & 3.57e-05 & 8.15 & 0.000156 & 0.44\\ 
  NormGraphlet-3-5 & 8.15 & 5.04e-05 & 6.69 & 0.00124 & 0.51\\ 
  NormOrderedGraphlet-3 & 10.50 & 4.33e-05 & 9.92 & 0.000135 & 0.39\\ 
  NormOrderedGraphlet-3-4 & 4.31 & 0.000999 & 4.92 & 0.00127 & 2.41\\ 
  NormOrderedGraphlet-3-4(K) & \textbf{1.69} & - & \textbf{2.08} & - & 2.41\\ 
  \hline
  GDDA & 17.30 & 6.16e-09 & 17.70 & 2.57e-08 & 0.54\\ 
  RGFD & 9.46 & 6.84e-06 & 9.85 & 1.39e-05 & 0.49\\ 
  GCD & 17.10 & 1.21e-09 & 17.10 & 1.51e-08 & 1.32\\ 
  GR-align & 8.31 & 0.00705 & 9.69 & 0.00423 & 9.49\\ 
  \hline
  Average degree & 18.90 & 2.32e-10 & 16.20 & 2.02e-07 & 0.39\\ 
  Average distance & 15.40 & 9.54e-07 & 16.50 & 3.59e-06 & 0.48\\ 
  Maximum distance & 17.30 & 1.58e-09 & 16.90 & 4.95e-08 & 0.49\\ 
  Average closeness centrality & 18.50 & 2.18e-08 & 16.50 & 3.08e-07 & 0.48\\ 
  Average clustering coefficient & 16.80 & 5.01e-08 & 14.50 & 3.55e-07 & 0.56\\ 
  Intra-hub connectivity & 16.40 & 2.84e-08 & 15.10 & 1.14e-06 & 0.64\\ 
  Assortativity & 20.10 & 1.79e-08 & 19.20 & 1.48e-07 & 0.46\\ 
  Existing-all & 10.90 & 1.33e-06 & 10.00 & 3.05e-05 & 1.01\\ 
  \hline
  DaliLite & 12.70 & 3.27e-05 & 10.60 & 0.00192 & 2021.41\\ 
  TM-align & 22.00 & 1.85e-12 & 22.30 & 5.75e-12 & 168.32\\ 
  Sequence & 14.50 & 1.44e-06 & 16.60 & 2.1e-08 & 0.24\\ 

\hline
\end{tabular}
\label{tab:running-time} 
%\end{center}
\end{SCtable}

Another of our key contributions is that even our regular graphlet PCA
approaches and especially their normalized and (``long-range($K$)'')
ordered counterparts are superior to traditional 3D contact
approaches, even though both approach types (network vs. 3D contact)
use 3D structural information. This highlights the usefulness of
network analyses of protein structures. This is especially true given
that our network approaches are also faster than the 3D contact
approaches, as follows.

\subsubsection{Running time comparison}
\label{results:real:running-time}

All alignment-free network approaches are comparable in terms of
running time to each other as well as to the sequence approach, they
are followed by the only alignment-based network approach (GR-align),
and all of them are significantly faster than the 3D contact
approaches (Table
\ref{tab:running-time}).

\subsubsection{Summary of results for PSNs}
\label{results:real:graphlet-vs-nongraphlet}

The non-normalized versions of our graphlet PCA approach are superior
to the existing graphlet (non-PCA), non-graphlet, 3D contact, and
sequence approaches. By normalizing the graphlet measures, we improve
upon the non-normalized measures and thus upon the existing methods
(Table \ref{tab:real-aupr} and Supplementary Table S3).
%\ref{supple:tab:real-auroc}).  
By imposing
sequence order onto nodes via ordered graphlets, we further improve
the accuracy. By distinguishing between shorter- and longer-range
amino acid interactions via ``long-range($K$)'' ordered graphlets, we
further improve the performance. NormOrderedGraphlet-3-4(K) is
superior to all considered methods in terms of its accuracy ranking
over all considered PSN sets, and its ranking is statistically
significantly better than the ranking of any other method (Table
\ref{tab:running-time}). This further validates our graphlet PCA
framework for PC.

\begin{figure}
  \begin{center} 
  \begin{minipage}[h]{0.75\linewidth}
  \raisebox{-3.5cm}{\epsfig{file=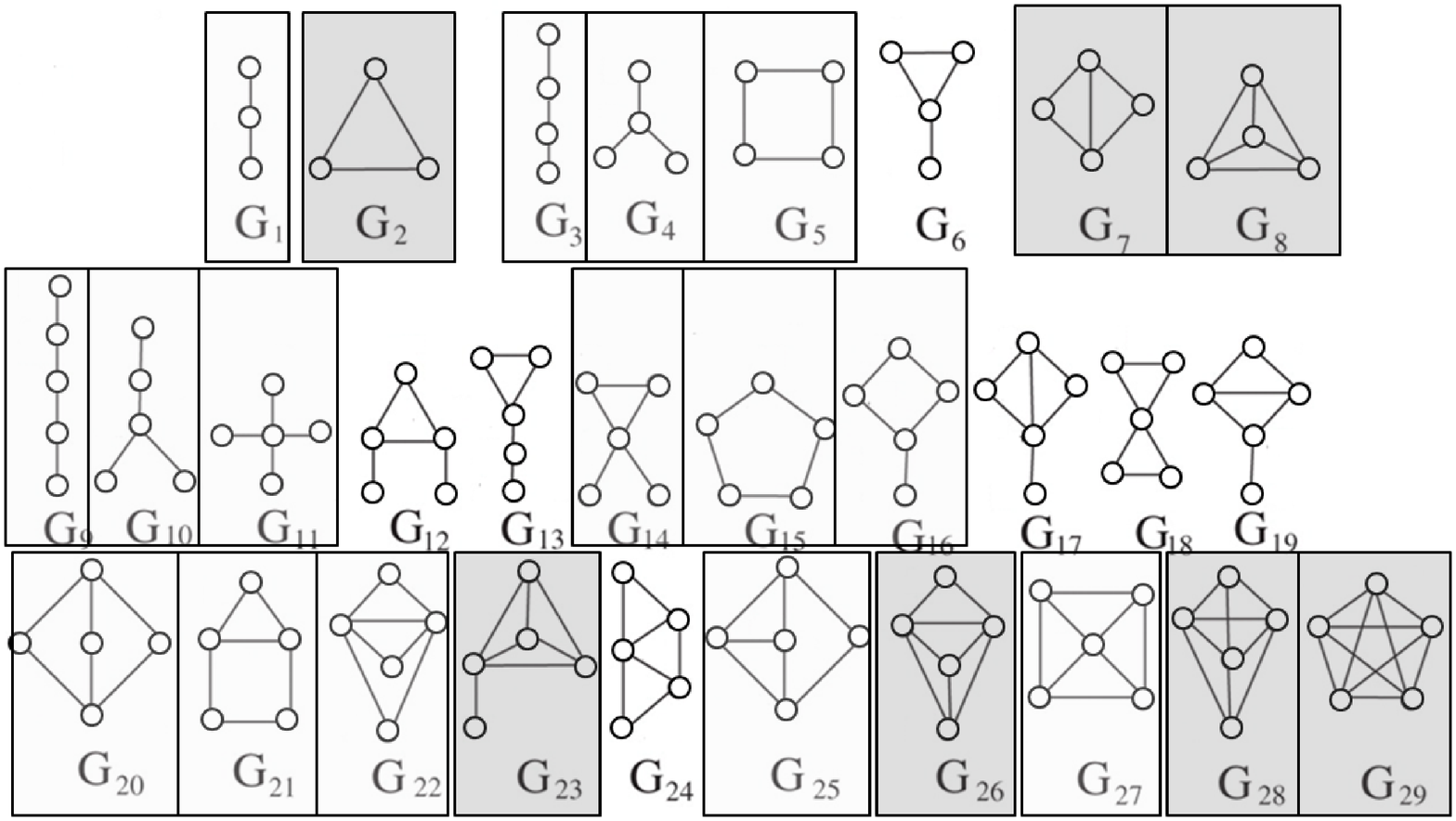,width=\linewidth}} 
  \end{minipage}\hfill
\begin{minipage}[h!]{0.24\linewidth} \caption{Regular (non-ordered) graphlets that are  significantly represented  in $\alpha$ (dark gray) or $\beta$ (light gray) PSNs. For equivalent results for ordered graphlets, see Supplementary Fig. S3.  %\ref{supple:fig:ordered-graphlet-cath-alpha-beta}. 
\label{fig:graphlet-cath-alpha-beta} } \end{minipage} \end{center}
\end{figure}

\subsection{Application of graphlet PCA measures in revealing 
biochemically interesting PSN patterns}
\label{results:application-graphlets}

We aim to identify graphlet patterns that lead to successful
distinction of different CATH or SCOP label categories from the PSN
data, focusing as an illustration on the PSN sets containing networks
of the same size (i.e., on CATH-95, CATH-99, and CATH-251-265) from
$\alpha$ or $\beta$ protein domain labels. Such graphlets that are
significantly (Mann-Whitney U test; $p < 0.05$) represented in
$\alpha$ but not in $\beta$, or vice versa, could be linked to the
functionality of the given domain label.

For the 3-5-node regular graphlet measure (i.e., Graphlet-3-5),
graphlets represented in $\alpha$ tend to be denser than those
represented in $\beta$ (Fig. \ref{fig:graphlet-cath-alpha-beta}).  For
example, all of the complete graphlets (i.e., $G_2, G_8, G_{29}$,
which are the densest graphlets) are represented in $\alpha$, while
all of the path-like graphlets (i.e., $G_1, G_3, G_9$, which are the
sparsest graphlets) are represented in $\beta$.

For the 3-4-node ordered graphlet measure (i.e., OrderedGraphlet-3-4),
in ordered graphlets represented in $\alpha$ (e.g., $O_1$), there is
typically a node order-respecting path through the graphlet, unlike in
most of ordered graphlets represented in $\beta$ (e.g., $O_2$ and
$O_3$) (Supplementary Fig. S3).
%\ref{supple:fig:ordered-graphlet-cath-alpha-beta}).  
Note that
for the data sets from this section (CATH-95, CATH-99, and
CATH-251-265), the ``long-range($K$)'' constraint does not improve
accuracy, and so we do not consider NormOrderedGraphlet-3-4($K$) here.

Linking the identified domain label-specific PSN patterns to their
potential biochemical meaning is our future interest.

\section{Conclusions}

We present a general computational framework for network comparion,
which can use any measure(s) of network topology.  We demonstrate the
effectiveness of our framework in the context of PC, in particular the
power of using graphlets as state-of-the-art network measures.
Specifically, we use ordered graphlets to integrate via network
analysis complementary protein 3D structural data and sequence data,
which improves upon the existing network (graphlet or non-graphlet),
3D contact, and sequence approaches. In the process, we address the
network size bias of the existing approaches.

%\clearpage
\bibliography{}
%\clearpage

\section*{Acknowledgements}

This work was supported by the National Science Foundation (CAREER
CCF-1452795 and CCF-1319469), Air Force Office of Scientific Research
(Young Investigator Program (YIP) FA9550-16-1-0147), National
Institutes of Health (1R01GM120733-01A1, 1R21AI111286-01A1, and R01
GM074807), and Clare Boothe Luce Graduate Research Fellowship.

\section*{Author contributions statement}

F.E.F., J.L.C., P.L.C., and T.M. designed the study. F.E.F. implemented all of the proposed methodology and carried out all of the experiments, with the following exceptions: K.N. implemented the ``long-range($K$)" approach within the proposed methodology and performed all experiments related to this approach, and J.L. suggested the use of PCA within the proposed methodology. J.L.C. assembled the protein structure datasets. F.E.F., K.N., J.L., S.J.E., P.L.C. and T.M. analyzed the results. F.E.F., J.L.C. K.N., P.L.C., and T.M. wrote the manuscript. All the authors read and approved the manuscript. P.L.C. supervised all applied aspects of the study. T.M. supervised all computational aspects of the study. 

\section*{Additional information}

\textbf{Competing financial interests:} The authors declare no competing financial interests.

\clearpage

%% Figures

%%Tables

\end{document}